\documentclass[preprint,showpacs,preprintnumbers,superscriptaddress,amsmath,amssymb]{revtex4-1}

\usepackage{graphicx}
\usepackage{dcolumn}
\usepackage{bm}

\begin{document}

\title{Local structural studies of Ba$_{1-x}$K$_x$Fe$_2$As$_2$ using atomic pair
distribution function analysis}
\author{B. Joseph}
\affiliation{Dipartimento di Chimica, INSTM (UdR Pavia), Universit\`{a} di Pavia, 
Viale Taramelli, 27100 Pavia, Italy}
\author{V. Zinth}
\affiliation{Department Chemie, Ludwig-Maximilians-Universit\"{a}t M\"{u}nchen, 
Butenandtstra. 5-13(Haus D), 81377 M\"{u}nchen, Germany }
\author{M. Brunelli}
\affiliation{Institut Laue-Langevin (ILL), 6 rue Jules Horowitz, BP 156, 38042 Grenoble Cedex 9, France}
\author{B. Maroni}
\affiliation{Dipartimento di Chimica, INSTM (UdR Pavia), Universit\`{a} di Pavia, 
Viale Taramelli, 27100 Pavia, Italy}
\author{D. Johrendt}
\affiliation{Department Chemie, Ludwig-Maximilians-Universit\"{a}t M\"{u}nchen, 
Butenandtstra. 5-13(Haus D), 81377 M\"{u}nchen, Germany }
\author{L. Malavasi}
\affiliation{Dipartimento di Chimica, INSTM (UdR Pavia), Universit\`{a} di Pavia, 
Viale Taramelli, 27100 Pavia, Italy}

\begin{abstract}
Systematic local structural studies of Ba$_{1-x}$K$_x$Fe$_2$As$_2$ system are undertaken at room temperature
using atomic pair distribution function (PDF) analysis. The local structure of the Ba$_{1-x}$K$_x$Fe$_2$As$_2$ 
is found to be well described by the long-range structure extracted from the diffraction
experiments, but with anisotropic atomic vibrations of the constituent atoms ($U_{11}$ = $U_{22} \ne U_{33}$).
The crystal unit cell parameters, the FeAs$_4$ tetrahedral angle and the pnictogen height above the
Fe-plane are seen to show systematic evolution with K doping, underlining the importance of
the structural changes, in addition to the charge doping, in determining the properties of
Ba$_{1-x}$K$_x$Fe$_2$As$_2$.

\end{abstract}


\maketitle

\section{Introduction}

The appearance of the iron-based superconductors \cite{kamihara}, have given a further boost to 
the efforts to understand the role of lattice elastic effects and disorder in layered 
structures \cite{dagatto} showing intriguing phenomena like high temperature 
superconductivity, colossal magneto-resistance etc.
The importance of local structural information in these complex systems is well recognised and
is the key driving force behind the development of experimental and theoretical tools associated
with the total scattering techniques \cite{pdf_book,pdf_review}. Several local structural 
studies in cuprates \cite{bianconi_prl96,billinge_prb93,bozin_prl2000}, 
manganites \cite{billinge_prl96, lanzara_prl98} cobaltites \cite{malavasi_prb2009} {\it etc}. 
have underlined the importance of the local order in the
functional properties of complex materials with hetero-structures. Polarized 
extended x-ray absorption fine structure (EXAFS) studies on
La$_{1.85}$Sr$_{0.15}$CuO$_{4 }$crystal have established two different conformations of 
the CuO$_6$ octahedra below 100 K \cite{bianconi_prl96}. In Nd$_{2-x}$Ce$_x$CuO$_{4-y}$ system, 
atomic pair distribution function (PDF) studies showed that the CuO$_{2}$ planes are not flat but are 
buckled and distorted by oxygen displacements of magnitude $\le$ 0.1 \AA $ $ \cite{bozin_prl2000}. 
PDF studies indicated that in local 
scale, the SmFeAsO$_{1-x}$F$_{x}$ compounds have a lower symmetry than the symmetry expected from the 
diffraction data measuring the long-range (average) structure \cite{pdf_sm1111}. EXAFS studies on the 
same system indicated the
presence of anomalies in the Fe-As bond correlation in the superconducting sample, with no
such anomalies being observed for the parent compound \cite{exafs_sm1111}. Even for the simplest 
systems among the new Fe based superconductor, {\it i.e.} the doped iron-chalcogenides, a lower symmetry is
observed at the local scale \cite{pdf_fesete}, with the presence of two distinct Fe-chalcogen 
bondlengths \cite{exafs_fesete,xrd_fesete}. To further understand the role of local structural 
inhomogeneities in the superconducting and magnetic properties, here we have undertaken the 
atomic pair distribution function analysis of the Ba$_{1-x}$K$_x$Fe$_2$As$_2$ system. 
Unlike the SmFeAsO$_{1-x}$F$_x$  \cite{pdf_sm1111} and 
the FeSe$_{1-x}$Te$_{x}$ \cite{pdf_fesete,exafs_fesete} systems, present results indicate that the 
local structure of Ba$_{1-x}$K$_x$Fe$_2$As$_2$ is well described by the
long-range (average) structure obtaineded from the diffraction experiments. However, the results of the 
PDF refinement suggest the atomic vibrations to be anisotropic, with the thermal fluctuations of the
constituent atoms along the $ab$ plane being not equal to the $c$ direction ($U_{11}$ = $U_{22} \ne U_{33}$). 
The crystal unit cell parameters, the Fe-As-Fe tetrahedral angle and the pnictogen height above the
Fe-plane are seen to show systematic evolution with K doping, underlining the importance of
the structural changes, in addition to the carrier supply, in determining the doping dependent
properties of Ba$_{1-x}$K$_x$Fe$_2$As$_2$.

\section{Experimental methods}

Ba$_{1-x}$K$_x$Fe$_2$As$_2$ (x=0.0, 0.1, 0.2 and 0.34) samples prepared via solid state reactions are used for
the present study. The synthesis, structural and transport properties of these samples are
reported in refs. \cite{bafe2as2_dis,bakfe2as2_dis, bakfe2as2_xrd}. X-ray total scattering measurements 
were carried out at the ID-31 beamline of the European synchrotron radiation facility, 
Grenoble (France). Finely powdered samples tightly packed in a 6 mm quartz capillary are used for 
the measurements. The measurements are conducted at temperature T = 297 K, using photons of 
wavelength 0.4008 \AA$~$(30 keV). As required for the atomic pair distribution function (PDF) 
analysis \cite{pdf_book}, data were collected with sufficient statistics to
high $Q$ values by measuring large angular range (2$\theta$ range 0 to 110 degrees). Data from an
empty container were also collected to take care of the container contributions. The corrected
total scattering structure function, $S(Q)$ is obtained using the standard corrections \cite{pdf_book} 
utilising the {\it PDFgetX2} program. From the $S(Q)$, the PDF data, $G(r)$, is obtained by the 
Fourier transformation according to 
$G(r) = \frac{2}{\pi} \int{Q[S(Q)-1]sin(Qr) dQ}$, 
where $Q$ is the magnitude of the scattering vector. Modelling of the PDF data is carried out 
using the {\it PDFgui} and {\it PDFfit2} packages \cite{pdfgui}.

\section{Results and discussions}

Figure 1(a) presents the room temperature atomic pair distribution function (PDF) data of the 
Ba$_{1-x}$K$_x$Fe$_2$As$_2$ ($x$=0.0, 0.1, 0.2 and 0.34) system. At room temperature the 
Ba$_{1-x}$K$_x$Fe$_2$As$_2$ system has a tetragonal symmetry, with the unit cell parameters 
of the parent compound ($x$=0.0) being $a = b \approx$ 3.96 \AA$~$ and 
$c \approx$ 13.0 \AA. With increasing $x$, $a$($b$) decreases and $c$ increases, with the latter 
having a larger effect, leading to an over-all expansion of the unit cell volume. However, the 
changes in the unit-cell volume are not so significant due to the similar ionic radii of 
Ba$^{2+}$ (1.42 \AA) and K$^+$ (1.51 \AA) for the same coordination. 
The overall PDF structure for all the samples studied are found to be similar
[Fig. 1(a)]. However, changes can be observed upon a careful comparison of the spectra
corresponding to different $x$ values. For example, the peak around 9 \AA$~$ becomes broader and
shows a splitting with increase in K doping. To obtain quantitative information, refinement of
the structure using the PDF data is undertaken. It is found that the average structure obtained
from the diffraction results gives a reasonable description of the PDF data. Anisotropic thermal
factors for the constituent atoms (according to the crystal symmetry constrains), rather than
isotropic, are found to result in a substantial improvement in the refinement. Refinement results
in such a case, for the $x$=0 and $x$=0.34 compounds are shown in Fig. 1(b).

In Fig. 2, left panels, we show the evolution of the lattice parameters, $a$($b$) and $c$, with 
K doping at room temeperature obtained from the PDF refinements. The data from the PDF refinement
have the similar trend as the diffraction data \cite{bakfe2as2_xrd}. The left upper inset in Fig. 2 
presents the changes in the $z$ position of the arsenic atom with K doping at room temperature. 
The $z$-position of the arsenic seems to be more or less constant (shows a small decrease with 
increasing K doping). It is interesting to follow the changes occuring in the active layer, 
Fe-As layer, in particular the FeAs$_4$ tetrahedra with the K-doping. The Fe-As distance show only 
little changes with K doping (Fig. 2, upper right panel), whereas the Fe-Fe distance is seen to 
show around 1\% decrease for the highest doped sample compared to the undoped. These changes 
result in a decrease in the Fe-As-Fe angle from around 110.82 to 109.48 [Fig. 2, inset in the 
upper right panel], leading to an increase in the pnictogen height above the Fe-plane 
with K-doping. In the iron-based superconductors, the anion height above the Fe-plane \cite{anion_mizuguchi}
is identified as a key parameter. Local structural studies on the {\it RE}FeAsO ({\it RE}=La, Pr, Nd and Sm) 
system showed that the Fe-As bondlength does not vary significantly with the change in the rare-earth {\it RE}
ionic size, but the Fe-Fe bondlength decrease with decreasing {\it RE} ionic size, thus 
affecting the arsenic height above the Fe-plane \cite{iadecola_REFeAsO}.
Several theoretical calculations reveal that the pnictogen height 
above Fe-plane is the controlling parameter of the magnetic interactions \cite{anion_height1,anion_height2}. 
Theoretical calculations indicate that even a small change in the arsenic-height above the 
iron-plane influences the density of states close to the Fermi-level \cite{anion_height1} and 
thus the magnetic and superconducting properties of the system. Substantial changes in the 
magnetic and superconducting properties from BaFe$_2$As$_2$ to Ba$_{0.66}$K$_{0.34}$Fe$_2$As$_2$ 
are in accordance with this. 

As mentioned previously, the PDF refinements are carried out considering anisotropic thermal
factors for the constituent atoms with ($U_{11}$ = $U_{22} \ne U_{33}$). In Fig. 3(a), we 
present the $U_{11}$ (= $U_{22}$) thermal factors for the constituent atoms. For the Ba site, 
the in-plane thermal factors seem to increase with increasing K doping. This can be understood 
as the K doping induced disorder in the Ba site, due to the slight differences in the ionic radii 
of the two. A similar increase is also seen for the Fe thermal factors. But the arsenic thermal 
factors seems to be more or less insensitive to the K doping. The thermal factors of the 
constiuent atoms along $c$ direction are given in Fig. 3(b). The $U_{33}$ for the Ba and Fe 
have larger magnitudes compared to the $ab$ plane thermal factors ($U_{11}$ = $U_{22}$).  
However, the $U_{33}$ for arsenic is lower in magnitude than the $ab$ plane thermal factor 
($U_{11}$ = $U_{22}$). In both cases ({\it i.e.} along $ab$ plane and along $c$), the 
As thermal factors are found to be lowest compared to the Ba/K and Fe.

In conclusion, we have carried out systematic local structural studies of Ba$_{1-x}$K$_x$Fe$_2$As$_2$ system
at room temperature using atomic pair distribution function (PDF) analysis. The local structure
refinements indicate that the Ba$_{1-x}$K$_x$Fe$_2$As$_2$ system does-not show any significant deviation 
in the structural symmetry at the local scale. However, for a better description of the local structure,
the assumption of anisotropic atomic vibrations of the constituent members are necessary, with
the thermal fluctuations along $ab$ plane and along $c$ having different magnitudes. Results also
show a systematic evolution of the FeAs$_4$ tetraheadra, with changing pnictogen height above
the Fe-plane with K doping, underlining the importance of the structural changes, in addition to
the charge-injection, in determining the properties of Ba$_{1-x}$K$_x$Fe$_2$As$_2$.

\begin{acknowledgments}
We acknowledge the European Synchrotron Radiation Facility for providing the beam time.
Research at {\it Universit\`{a} di Pavia} is supported by the {\it CARIPLO} foundation (Project No. 
2009-2540 {\it Chemical Control and Doping Effects in Pnictide High-temperature Superconductors}).
\end{acknowledgments}


\newpage

\begin{figure}
\includegraphics[width=12.0cm]{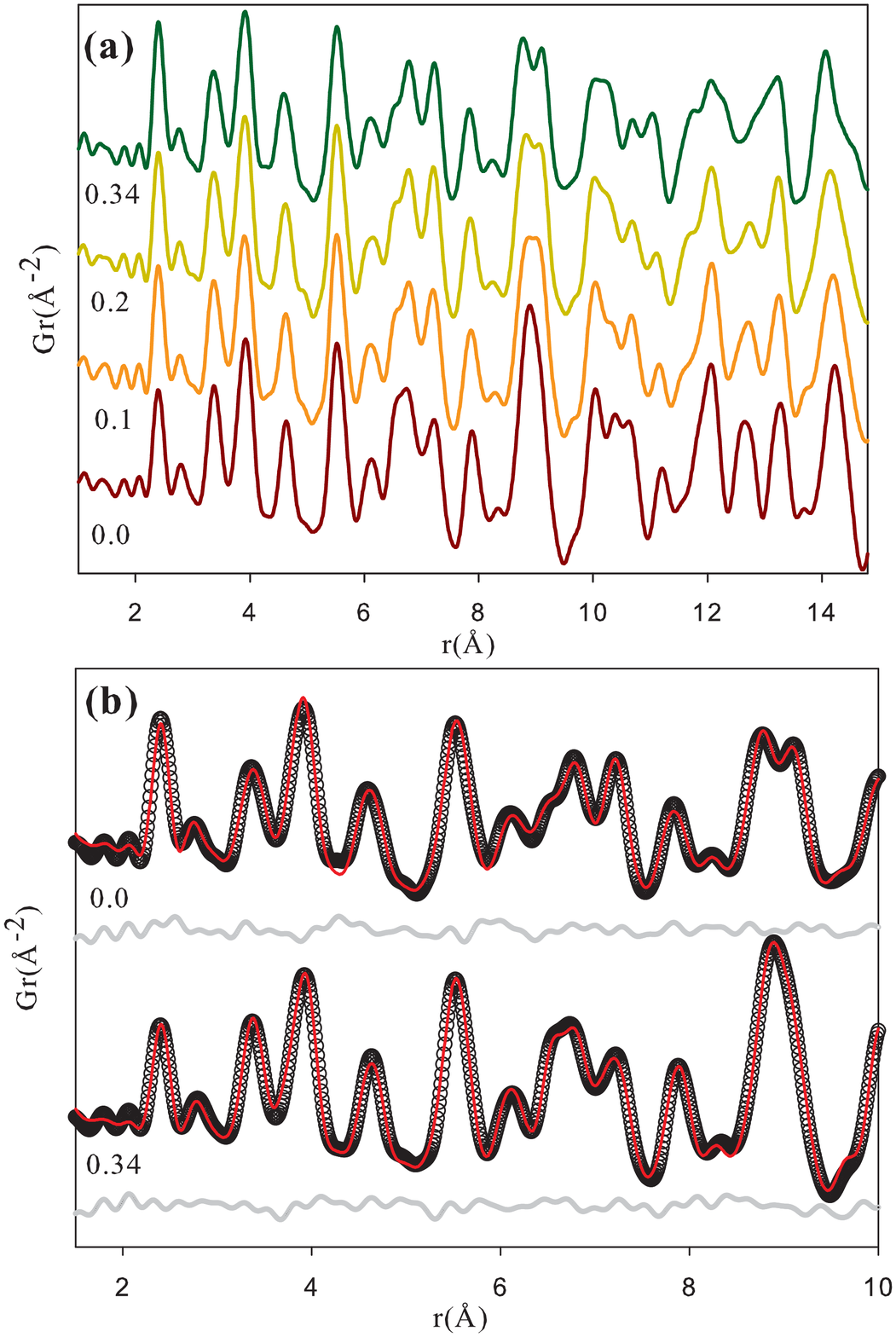}
\caption{\label{fig:epsart} 
(a): Shows the PDF data at room temperature for the Ba$_{1-x}$K$_x$Fe$_2$As$_2$
($x$=0.0, 0.1, 0.2 and 0.34) system. (b): Refinement results for the BaFe$_2$As$_2$ and
Ba$_{0.66}$K$_{0.34}$Fe$_2$As$_2$ samples. The refinements were done taking the starting parameters 
from the diffraction studies with anisotropic thermal factors (according to the space group constrains).
}
\end{figure}

\newpage

\begin{figure}
\includegraphics[width=15.0cm]{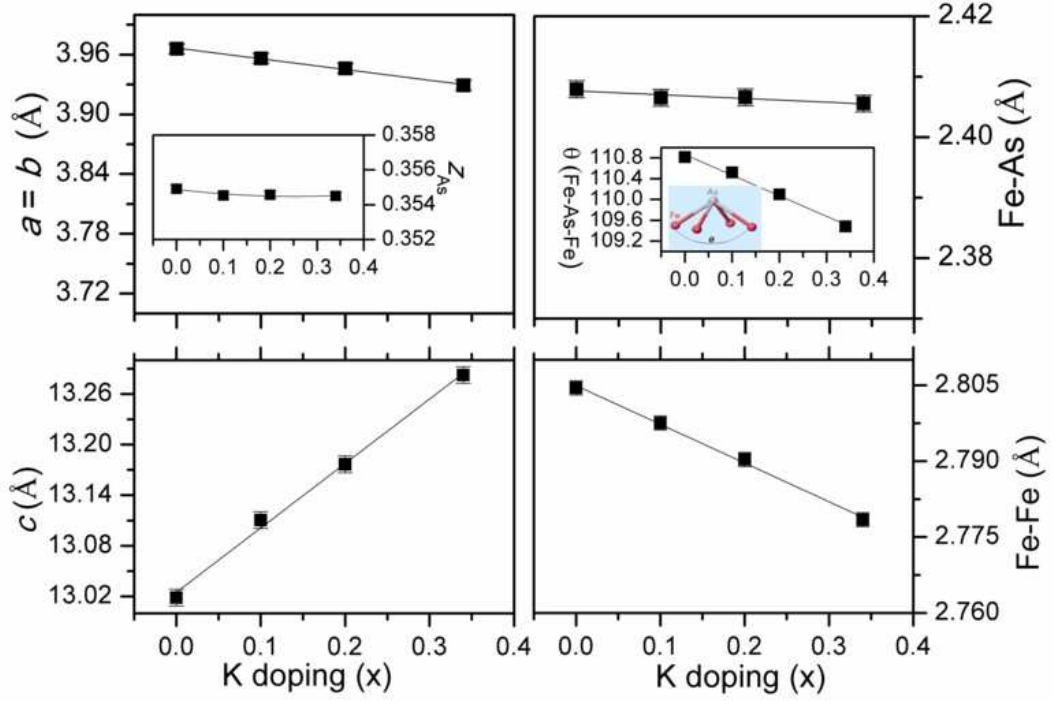}
\caption{\label{fig:epsart}
Left panel: K doping dependent lattice parameter [($a$=$b$) and ($c$)] variations extracted
from the refinement of the Ba$_{1-x}$K$_x$Fe$_2$As$_2$ structure using PDF data at room 
temperature. Inset in the upper panel shows the K doping dependenence of the $z$-position of 
the As atom. Right panel: Doping dependence of the Fe-As and Fe-Fe distances extracted from 
the room temperature refinement of the Ba$_{1-x}$K$_x$Fe$_2$As$_2$ structrue using the 
PDF data. Inset in the upper panel shows the variation of the Fe-As-Fe angle. Dotted line 
in all panels is only a guide to the eyes.}
\end{figure}

\newpage

\begin{figure}
\includegraphics[width=12.0cm]{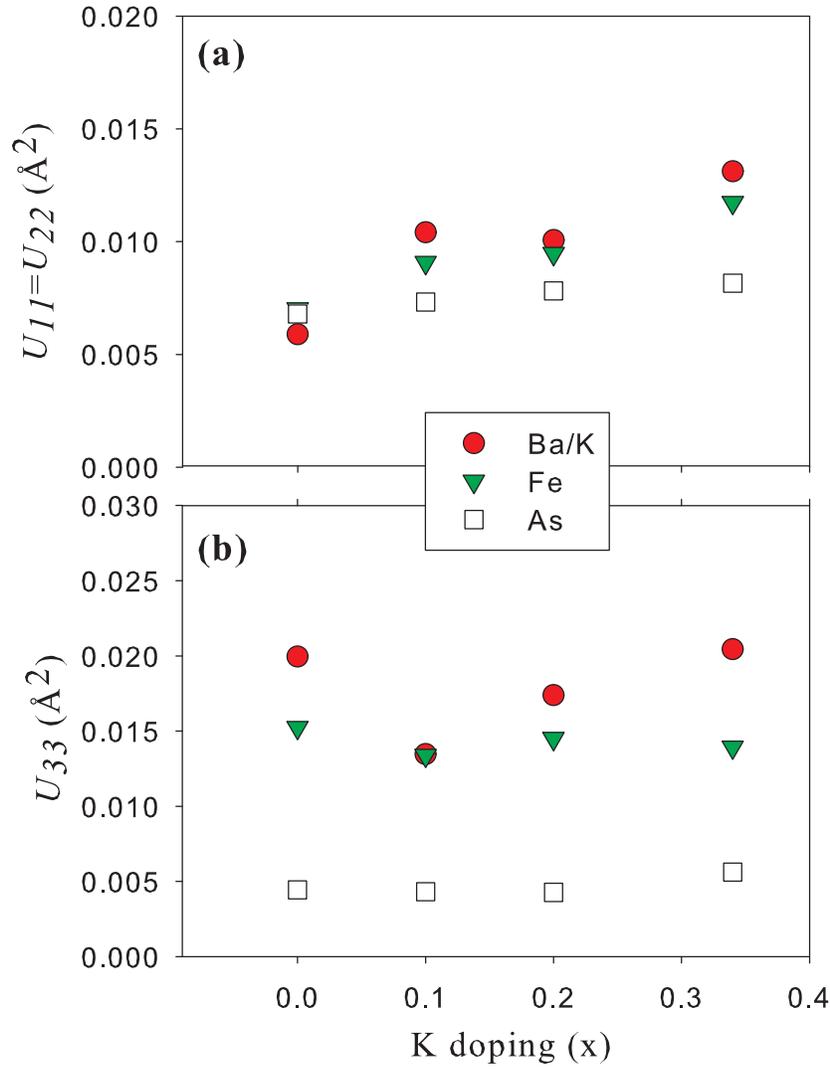}
\caption{\label{fig:epsart} 
(a): Doping dependenence of thermal factors of Ba, Fe and As along the $ab$ plane
($U_{11}$=$U_{22}$) extracted from the room temperature refinement of the Ba$_{1-x}$K$_x$Fe$_2$As$_2$ 
structure using the PDF data. (b): Doping dependenence of thermal factors of Ba, Fe and As 
along $c$ ($U_{33}$) extracted from the room temperature refinement of the Ba$_{1-x}$K$_x$Fe$_2$As$_2$ 
structure using the PDF data.
}
\end{figure}

\end{document}